\documentclass[12pt]{article}
\usepackage{graphicx}

\begin{document}

\begin{center}
\LARGE Tau polarization in quasielastic charged-current
neutrino(antineutrino)-nucleus scattering
\\[4mm]
\end{center}

\begin{center}
\large Krzysztof M. Graczyk\footnote{kgraczyk@ift.uni.wroc.pl}
\\[2mm]
July 23, 2004
\end{center}

\begin{center} Institute of Theoretical Physics,
University of Wroc\l aw              \\
pl. M. Borna 9, 50 -- 204 Wroc\l aw, Poland \\
Phone:(+48-71) 375-9408 \\
Fax: (+48-71) 321-4454 \end{center}

\begin{footnotesize}\textbf{PACS codes}: 13.15+g, 25.30Pt, 13.88+e

\end{footnotesize}


 \abstract{ The quasielastic charged-current (CC) tau neutrino
 (antineutrino)-nucleus scattering is considered.
 The dependence of tau polarization on nuclear-structure effects is discussed in detail.
 The description of the nucleus is based on the Mean-Field Theory (MFT).
 The ground state of nucleus is described using  the Relativistic Fermi Gas model (FG).
  The effective mass is introduced as well as the ring Random Phase Approximation
  (RPA) effects are taken into account in the framework
  of relativistic meson-nucleon model.  The Local Density Approximation (LDA) is
  used for the argon nucleus, having in mind possible application to the ICARUS experiment.
  The discussion concentrates on the threshold region where  the $\tau^-$ can be unpolarized and
 the nuclear effects play an important role. }

\section{Introduction}
The oscillation of $\nu_\mu$ into $\nu_\tau$ is a possible
explanation of the deficit of the atmospheric muon neutrinos which
has been  measured in SuperKAMIOKANDE. New projects such as ICARUS
and OPERA  will enable a more precise study of the $\nu_\mu
\rightarrow\nu_\tau$ oscillations phenomena by the detection of
tau neutrinos. In the case of the CNGS beam and the ICARUS
detector with five T600 modules  about 12 events of $\nu_\tau$ are
expected (after kinematical selection procedure) during five years
of data taking (assuming $\delta m^2= 2.5 \times 10^{-3} eV^2$
)~\cite{hiszpanka}. In the case of the OPERA 1.8kT detector about
11 events are expected. This small statistic requires detailed
analysis which should be done based on precise theoretical
predictions for the considered process. In the case of tau
neutrino-matter interaction the produced lepton has short
lifetime, so only its decay products can be observed. The large
mass of the tau in contrast to electron and muon implies that it
can be partially polarized~\cite{Smith,Albright}. The degree of
polarization of tau is one of the parameters which describe its
decay products distributions~\cite{Okun,Leader}. Therefore, the
discussion the tau polarization in the neutrino-matter scattering
can play an important role in the analysis of the experimental
data and is also of theoretical interest.

Polarization properties of taus produced in charged-current
neutrino-matter interactions have been considered by several
groups~\cite{japonczycy,rosjanie,Soffer}. In
papers~\cite{japonczycy,rosjanie} the polarization vector of the
tau was calculated by using the spin density
matrix~\cite{Landau,Leader}. Authors discuss neutrino-matter
interaction in the  three regions namely quasielastic
neutrino-nucleon scattering, neutrino-nucleon scattering with
single pion production and deep inelastic scattering. As was shown
by K. Hagiwara  et al.~\cite{japonczycy} the produced leptons are
characterized by high polarization degree.  In the laboratory
frame (LAB) the low energy $\tau^-$s can be right-handed (positive
sign of the longitudinal polarization) and the  high energy
$\tau^-$s are always left-handed (have negative helicity). For
$\tau^+$ there exists a similar effect. This effect can be
explained by considering the $\nu-n$ scattering in the center of
mass  frame (CM) where the produced $\tau$s can be scattered in
both forward and backward directions. In the CM frame the produced
$\tau^-$ is always left-handed (helicity has negative sign).
However, the left-handed taus which are scattered in backward
directions in the CM frame usually become right-handed and are
scattered in the forward directions after performing the boost to
the LAB frame.

A covariant way of calculation of the spin density matrix was
presented by K.S. Kuzmin et al. in~\cite{rosjanie}. In the
description of the single pion production they used the
Rein-Sehgal model. They also considered quasielastic
neutrino-nucleon scattering using nonstandard vector and axial
currents.

The aim of this paper is the investigation of  how the nuclear
matter affects the tau polarization.

The relativistic mean-field theory formalism is used to describe a
nucleus \cite{Walecka}. This description   is based on
relativistic Fermi gas (FG) model~\cite{Smith_Moniz}. The nucleons
inside the nucleus do not interact with each other. Their momenta
are uniformly distributed in the Fermi sphere which of the radius
given by the Fermi momentum  $k_F$. There is a direct relation
between the $k_F$ and the nuclear matter density.

To make the description of the nucleus more realistic the
effective mass of nucleons is considered and the ring random phase
approximation (RPA) is done based on the residual interaction
$\pi+\rho+g'$~\cite{Horowitz}. The RPA corrections are calculated
by taking into account the infinite sum of one particle -- one
hole diagrams (p-h). The effective mass is introduced following
Walecka ($\sigma$--$\omega$ model)   \cite{Walecka}. In the
simplest version of the model the mean-field approximation leads
to the modifications of the nucleon four-momentum and its mass (in
the fermion propagator). One of the main results of the mean-field
approximation of the $\sigma-\omega$ model  is a self-consistency
equation relating the effective mass with the Fermi momentum.
Solving this equation for $k_F=$225~MeV yields the effective mass
equal to 638~MeV. The effective mass is then introduced into the
fermion propagator instead of the free mass ($M\rightarrow M^*$).

Having in mind application of calculations to the ICARUS
experiment we discuss also neutrino-argon scattering. The argon
nucleus is described in the local density approximation
(LDA)~\cite{Oset}. The Fermi momenta of nucleons are local and
given by the experimental charge density
distributions~\cite{Moniz-Whitney}.

The paper is organized as follows. In the first part the degree of
polarization and the technical details of the description of the
nucleus model are given. The calculation of the polarization
vector is based on the algebraic decomposition of the nuclear
polarization tensor~\cite{Graczyk}. In the second part of the
article the numerical results are presented and discussed. In
general we get results similar to those of K. Hagiwara et
al.~\cite{japonczycy}. The tau leptons  are characterized by  high
degree of polarization for almost all energies and scattering
angles with the exception for the zero scattering angle where the
$\tau^-$ can be partially polarized.

In the discussion of the nuclear-structure effects we  focus on
how the polarization of the tau is affected by introducing to the
Fermi gas model the effective mass of the nucleon and the RPA
corrections. We show that in the case of forward scattering (zero
scattering angle) the mean value of degree of polarization of
$\tau^-$  has a minimum around the neutrino energy of 4.5~GeV. The
minimal value of $\left<\mathcal{P}_{\theta=0,E}\right>$ for the
basic FG model is 0.2 so the tau is almost unpolarized. The
introduction of the effective mass and the RPA corrections
increases this minimum to about 0.4.

\section{Theoretical description }

\subsection{Polarization}

We focus on the following  quasielastic processes:
\begin{eqnarray*}
\nu_\tau(k) + n(p) & \rightarrow & \tau^-(k',s^\mu) + p(p'), \\
\overline{\nu}_\tau(k) + p(p) & \rightarrow & \tau^+(k',s^\mu) +
n(p').
\end{eqnarray*}
The produced $\tau^-$ and $\tau^+$ are  polarized, its spin vector
$s_\mu$ satisfies (in any frame) the relations:
$$
{k'}_\mu s^\mu =0, \;\;\;\;\; s_\mu^2 = -1.
$$
The differential cross section is expressed by the following sum
of two contributions:
$$
d\sigma(k,q,s)  \sim  \left(\sum_\lambda \overline{\sum_s}
L_{\mu\nu}(\lambda,s)  + \sum_\lambda L_{\mu\nu}(\lambda,s)
\right)W^{\mu\nu}.
$$
The hadron tensor is defined as:
\begin{equation}
W^{\mu\nu}=\sum_{f} (2\pi)^4 \delta^4(q -{p_f} + {p_i})
\left<i|J^\mu(0)|f\right>\left<f|J^\nu(0)|i\right>,
\end{equation}
where $J_\mu$ -- electroweak current, $\left|i\right>$,
$\left|f\right>$ -- initial and final hadronic states. The
$L^{\mu\nu} (\lambda,s)$ denotes the lepton tensor, $\lambda$ --
neutrino helicity.

Summing over helicities leads to the expressions:
$$
\sum_\lambda\overline{\sum_s} L_{\mu\nu}(\lambda,s) = 8
\left(k_\mu {k'}_\nu +k_\nu {k'}_\mu -g_{\mu\nu}{k'}_\alpha
k^\alpha  \mp \mathrm{i}\epsilon_{\mu\nu\alpha\beta}k^\alpha
{k'}^\beta \right) \equiv L^0_{\mu\nu},
$$
$$
 \sum_\lambda L_{\mu\nu}(\lambda,s)  =   8 m s^\alpha \left(k_\nu g_{\alpha\mu}
+ k_\mu g_{\nu\alpha} - g_{\mu\nu} k_\alpha \mp
\mathrm{i}\epsilon_{\mu\nu\beta\alpha}k^\beta\right)\equiv
L^s_{\mu\nu},
$$
where sign $\mp$ corresponds to scattering of
neutrino/antineuytrino. The second expression is linear in the
lepton mass $m$ and the spin four-vector $s_\mu$. That is why in
the case of electron and muon the contribution to the cross
section due to polarization is most often neglected.

The polarization of the tau lepton measured in the direction of
the four-vector $s_\mu$ is given by the formula~\cite{Bjorken}:
\begin{equation}
\mathcal{P}_{s_\mu}   =  \frac{ d\sigma(k,q,s) - d\sigma(k,q,-s)}
{d\sigma(k,q,s) +d\sigma(k,q,-s)} = \frac{L_{\mu\nu}^s
W^{\mu\nu}}{L_{\mu\nu}^0 W^{\mu\nu}}=P_\mu s^\mu,
\end{equation}
which defines $P_\mu$ -- the polarization vector of the tau
lepton.

\nopagebreak We introduce the four-vectors $e_l^\mu$, $e_t^\mu$,
$e_p^\mu$~\cite{Smith} such that:
\begin{itemize}
 \item they are orthogonal and in the rest frame of the tau
$$
 e_l = ( 0, \mathbf{e}_l),\quad
 e_t=  (0,  \mathbf{e}_t),\quad
 e_p = (0,\mathbf{e}_p), \quad
|\mathbf{e}_l|=|\mathbf{e}_p|=|\mathbf{e}_t|=1;
$$
 \item in the laboratory frame their spacelike parts satisfy:
$$
 \mathbf{e}_l \sim \mathbf{k'},\quad
\mathbf{e}_t \sim \mathbf{k} \times \mathbf{k'},\quad
 \mathbf{e}_p \sim \mathbf{e}_l \times \mathbf{e}_t.
$$
\end{itemize}
It follows that in the laboratory frame have the form:
$$
e_l^\mu = \frac{1}{m}\left( |\mathbf{k'}|,E_\tau
\frac{\mathbf{k'}}{|\mathbf{k'}|} \right),\quad e_t^\mu=(0,
\mathbf{e}_t),\quad e_p^\mu = (0,\mathbf{e}_p).
$$

The decomposition of polarization four-vector in the  above basis:
\begin{equation}
P^\mu =  \alpha {k'}^\mu + e_l^\mu \mathcal{P}_l + e_p^\mu
\mathcal{P}_p +e_t^\mu \mathcal{P}_t
\end{equation}
defines its longitudinal $\mathcal{P}_l$, perpendicular
$\mathcal{P}_p$ and transverse $\mathcal{P}_t$ components.

To define the degree of polarization it is useful to go into the
rest frame of the tau, where the four-vector $s_\mu$ is spacelike
$$
s_\mu = (0,\mathbf{\hat{s}}), \;\;\; \mathbf{\hat{s}}^2 =1.
$$
Thus the polarization which is measured in the direction of
$\mathbf{\hat{s}}$ is equal to:
$$
\mathcal{P}_\mathbf{s} =-\mathbf{P}\cdot\mathbf{s} = -|\mathbf{P}|
\cos(\beta),
$$
 $\beta$ being the angle between $\mathbf{P}$ and
$\mathbf{\hat{s}}$.

The quantity
\begin{equation}
\label{Polaryzacja_def} \mathcal{P} \equiv |\mathbf{P}|
\end{equation}
is called the degree of polarization.

Because  $\mathbf{\hat{s}}$ is spanned  (in the rest frame of
$\tau$) by $\mathbf{e}_l,\mathbf{e}_t,\mathbf{e}_p$, hence the
degree of polarization is given by the longitudinal, transverse
and perpendicular polarizations. The Lorentz boost does not change
this properties in any frame:
\begin{equation}
\mathcal{P} =
\sqrt{\mathcal{P}_l^2+\mathcal{P}_p^2+\mathcal{P}_t^2}.
\end{equation}

Further consideration will be performed in the laboratory frame
because the description of the nucleus (in particular
distributions of the Fermi momenta) is established in the LAB
frame.

\subsection{Model of the nucleus}

In the model of the nucleus adopted in this paper the crucial role
is played by the polarization tensor $\Pi^{\mu\nu}$ which is
directly related to the hadronic tensor  $W^{\mu\nu}$. The
inclusive cross section (normalized per one nucleon) for the
neutrino-nucleus scattering is proportional to the imaginary part
of the contraction of the polarization tensor with the lepton
tensor~\cite{Horowitz}:
\begin{equation}
\label{przekroj} \frac{d^2 \sigma }{d \cos(\theta) \, d E_\tau}=
-\frac{ G_F^2 \mathrm{cos}^2\theta_c \; |\mathbf{k'}|}{16 \pi^2
\rho_F E } \mathrm{Im} \left ( {L_\mu}^\nu {\Pi_\nu}^\mu \right ).
\end{equation}
Here $\rho_F=k^3_F/3\pi^2$ is neutron or proton density in
nucleus. E is the neutrino(antineutrino) energy and $\theta$ --
the scattering angle. We fix the Fermi momentum at 225~MeV in the
presented calculations.

In general, the polarization tensor is defined by the
chronological product of many body currents:
$$
\Pi^{\mu\nu}(q_\alpha) = -\mathrm{i}\int d^4 x
e^{\mathrm{i}q_\alpha x^\alpha}
\left<0\right|T\left(\mathcal{J}^\mu(x)\mathcal{J}^\nu(0)\right)\left|0\right>.
$$
Usually to simplify the problem a one body current is considered
instead of a many body one. In the case of the charged-current
electroweak interaction the one body current is given by:
\begin{eqnarray}
\mathcal{J}^\mu(x) & = & \!\!\!\!\!\! \overline{\Psi}(x)\Gamma^\mu\Psi(x), \nonumber \\
\label{Gamma_mu} \Gamma^\alpha(q_\mu) & = & \!\!\!\!\!\!
F_1(q^2_\mu)\gamma^\alpha
+F_2(q_\mu^2)\frac{\mathrm{i}\sigma^{\alpha\nu}q_\nu}{2M} +
G_A(q_\mu^2)\gamma^\alpha\gamma^5 + F_p(q^2_\mu) q^\alpha
\gamma^5.
\end{eqnarray}
where the form factors $F_1$, $F_2$, $G_A$, and $F_p$ have well
known dipole form and are described by the following set of
parameters~\cite{Bodek}: $M_A=$1.0~GeV, $g_A=-1.26$, $\mu = 4.71$,
$M_V^2$=0.71~GeV$^2$.

The ground state of the nucleus is the relativistic Fermi gas. One
can easily obtain the polarization tensor in this case:
\begin{equation} \Pi_{free}^{\mu\nu}(q_\alpha) = -\mathrm{i}
\int \frac{d^4 p}{( 2 \pi )^4} \mathrm{Tr} \left ( G(p +
q)\Gamma^\mu(q)G(p)\Gamma^\nu (-q ) \right ),
\end{equation}
where $G(p)$ is  the nucleon propagator  in the Fermi sea.
\begin{equation}
 G(p) = (p \!\!\!\! / + M ) \left ( \frac{ 1 }{
p_\alpha^2 - {M}^2 + \mathrm{i}\epsilon } +
\frac{\mathrm{i}\pi}{E_p}\delta(p_0 - E_p)\theta(k_F - p) \right )
\end{equation}
As was mentioned, the mean-field approximation in the
$\sigma-\omega$ model leads to the modifications of nucleon
four-momentum and its mass which become effective
\cite{wehrberger}:
$$
p^*_\mu = (p_0^*-g_V^N V_0, \mathbf{p}), \quad M \rightarrow M^*,
$$
where $g_V^N$ is the coupling constant for the $\omega$-nucleon
interaction. The effective mass is given by the self-consistency
equation. In the simplest approach the potential $V_0$  is
constant and can be eliminated by the change of variables in the
integral which gives $\Pi^{\mu\nu}$. Hence only the effective mass
appears in further calculations.

The considered residual interaction $\pi+\rho+g'$  is described by
the effective lagrangian~\cite{Walecka} which is given by
nucleon-meson $\rho$ and nucleon-pion interaction terms. Following
H. Kim et al.~\cite{Horowitz} a short-range correlation effect is
taken into account by introducing the Landau-Migdal parameter
($g'=0.7$) into the pion propagator. Using the interaction
introduced above the ring random phase approximation is
performed~\cite{Fetter}. It is calculated by taking into account
the infinite sum of contributions from ring diagrams (one particle
-- one hole excitations). It leads to the correction of the free
polarization  tensor:
\begin{equation}
\Pi^{\mu\nu} = \Pi^{\mu\nu}_{free} + \Delta\Pi_{RPA}^{\mu\nu}.
\end{equation}
To simplify the analysis, it is convenient  to choose the
coordinate system which in  the four-momentum  transfer $q_\mu$
has the form $(q_0,q,0,0)$. Then the polarization tensor as a
matrix can be decomposed in four independent parts:
\begin{eqnarray}
{\Pi_{\mu}}^\nu = e_L\Pi^L + e_T \Pi^T+ e_{VA}\Pi^{VA} + e_A\Pi^A,
\end{eqnarray}
where  $\Pi_{L,T,A,VA}$  are complex functions which are called
longitudinal, transverse, mixed and axial.  $e_{L,T,A,VA}$ are
$4\times4$ matrices which form a closed algebra~\cite{Graczyk}.
Properties of this algebra allow us to obtain analytical solution
for each component of polarization tensor separately:
$$
\Pi_{L,T,A,VA} \rightarrow  \Pi_{L,T,A,VA} + \Delta
\Pi_{L,T,A,VA}^{RPA}.
$$
In presented calculations only density dependent part of the
polarization tensors is taken into account~\cite{Horowitz}.
Contributions from the divergent Feynman parts are omitted.

Decomposition of  the $\Pi^{\mu\nu}$ leads to the following
decomposition of the scattering amplitude:
\begin{equation}
L_{\mu\nu}\Pi^{\mu\nu} = L_L\Pi^L + L_T \Pi^T \pm L_{VA}\Pi^{VA} +
L_A\Pi^A,
\end{equation}
where:
\begin{eqnarray*}
L_L & \equiv& L_{\mu}^{\;\;\nu}{e_L}_\nu^{\;\mu} =
-\frac{q^2}{q^2_\mu}L_{00} + \frac{q_0
q}{q^2_\mu}(L_{01}+L_{10}) - \frac{q_0^2}{q^2_\mu}L_{11},\\
L_T & \equiv & L_{\mu}^{\;\;\nu}{e_T}_\nu^{\;\mu} =-(L_{22}+L_{33}),\\
L_{VA} & \equiv &L_{\mu}^{\;\;\nu}{e_{VA}}_\nu^{\;\mu} = 2 \mathrm{i}L_{23},\\
L_A & \equiv & L_{\mu}^{\;\;\nu}{e_A}_\nu^{\;\mu} ={L_{\mu}}^\mu
\end{eqnarray*}
are longitudinal, transverse, mixed and axial components of lepton
tensor.

As was shown, calculations of the degree of polarization requires
the knowledge of the longitudinal, transverse and perpendicular
components of polarization:
\begin{eqnarray}
\mathcal{P}_{i_\mu} & = &
\frac{d\sigma(E_\tau,\theta,i_\mu)-d\sigma(E_\tau,\theta,-i_\mu)
}{d\sigma(E_\tau,\theta,i_\mu)+d\sigma(E_\tau,\theta,-i_\mu)}=
\frac{\mathrm{Im}\left(L_{\mu\nu}^{i}
\Pi^{\mu\nu}\right)}{\mathrm{Im}\left( L_{\mu\nu}^0
\Pi^{\mu\nu}\right)} \\
& = & \frac{\mathrm{Im}\left(L_L^i\Pi^L + L_T^i \Pi^T \pm
L_{VA}^i\Pi^{VA} + L_A^i\Pi^A\right)}{\mathrm{Im}\left(L_L^0\Pi^L
+ L_T^0 \Pi^T \pm L_{VA}^0\Pi^{VA} + L_A^0\Pi^A\right)},
\end{eqnarray} where $i_\mu = e^l_\mu,e^t_\mu, e^p_\mu$.

Performing the corresponding decomposition of the lepton tensor we
obtain:
\begin{itemize}
\item longitudinal:
\begin{eqnarray*}
L_L^{l}    & = & \frac{8m}{q^2_\mu}\left( q^2(q_0 e_0^l - q e_1^l
- 2 E e_0^l ) +2
q_0 q (E e_1^l +k_1 e_0^l) \right)\\
&  & -\frac{8 q_0^2 m}{q^2_\mu}\left(q_0 e_0^l - q e_1^l +2k_1 e_1^l\right),\\
L_T^{l}    & = & -16 m \left(E e_0^l - k_1 e_1^l\right), \\
L_A^{l}    & = & -16 m (q_0 e_0^l - q e_1^l), \\
L_{VA}^{l} & = & 16 m  \left( E e_1^l -k_1\ e_0^l\right);
\end{eqnarray*}
\item perpendicular:
\begin{eqnarray*}
L_L^{p} & = & \frac{8m e_1^p}{q^2_\mu}\left( 2E q_0 q - q^3 +
q_0^2
q - 2q_0^2 k_1  \right),\\
L_T^{p} & = &  16 m e_1^p k_1, \\
L_A^{p} & = &  16 m q e_1^p, \\
L_{VA}^{p} & = & 16 m E e_1^p.
\end{eqnarray*}
\end{itemize}
The contribution from the transverse polarization vanishes which
means that the polarization vector lies in the scattering plane.
One can notice  that $\mathcal{P}_p$ vanishes  for zero scattering
angle because it is proportional to  $\sin(\theta)$. In the zero
lepton mass limit we get:
$$
(\mathcal{P}_l,\mathcal{P}_t,\mathcal{P}_p) \rightarrow (\mp
1,0,0),
$$
where sign $+$ ($-$) corresponds to right-handed, and left-handed
lepton helicity respectively.

It can be shown that the components of lepton tensor satisfy the
following "triangle relations":
\begin{eqnarray}
(L_L^l)^2 + (L_L^p)^2  & = & (L_L^0)^2, \\
(L_A^l)^2 + (L_A^p)^2  & = & (L_A^0)^2,
\end{eqnarray}
\begin{eqnarray}
(L_T^0)^2 - (L_T^l)^2 - (L_T^p)^2 & = &  (L_{VA}^l)^2 + (L_{VA}^p)^2 -(L_{VA}^0)^2,
\nonumber \\
&  &
\!\!\!\!\!\!\!\!\!\!\!\!\!\!\!\!\!\!\!\!\!\!\!\!\!\!\!\!\!\!\!\!\!\!\!
\!\!\!\!\!\!\!\!\!\!\!\!\!\!\!\!\!\!\!\!\!\!\!\!\!\!\!\!\!\!\!\!\!\!\!
=-\frac{64m^2}{q^2}\left(4 q_\mu^2 E_\tau  E  + 4(E q_0 - q_\mu^2
)m^2 +( m^2 +q_\mu^2 )^2 \right),
\end{eqnarray}
which make  numerical calculations  easier.

\section{Numerical results and discussion}

\begin{figure}[p]
\centerline{
\includegraphics[width=18cm]{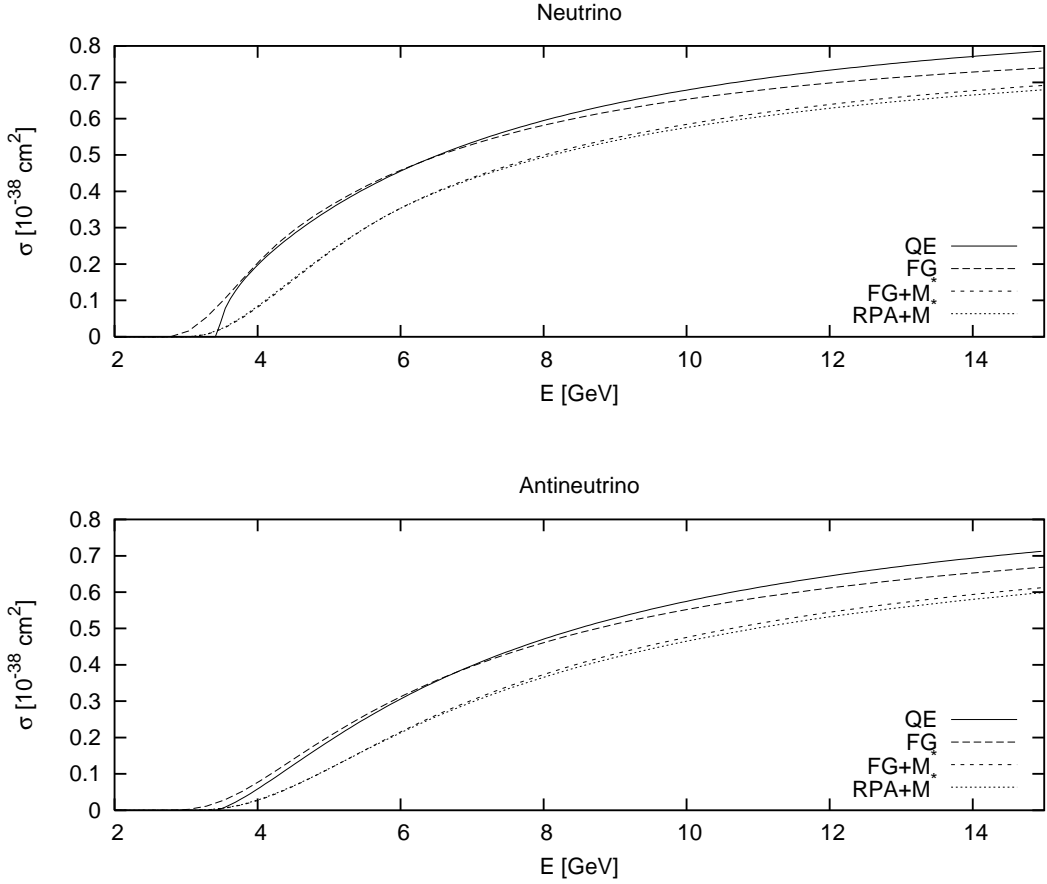}
} \caption{Comparison of the total cross sections (per one
nucleon) for charged-current quasielastic neutrino and
antineutrino-nucleus scattering. The model of nucleus is the Fermi
gas (dashed line -- FG), the Fermi gas with the effective mass
$M^*$=638~MeV (short dashed line -- FG+$M^*$) and the Fermi gas
with the effective mass  $M^*$=638~MeV and the RPA corrections
(dotted line -- RPA+$M^*$). The  total cross sections calculated
for neutrino and antineutrino-free nucleon scattering are shown
(solid line -- QE). \label{wykres_pol1} }
\end{figure}

We begin the discussion of the numerical results with the
presentation of the total cross sections. In the
Fig.~\ref{wykres_pol1} the cross sections for quasielastic
scattering on nucleus for both neutrino and antineutrino are
compared to the scattering on a free nucleon. The scattering on
the nucleus is described by the Fermi gas model. The Pauli
blocking reduces the total cross section by about 8\% compared to
the free nucleon case while the Fermi motion shifts the cutoff
energy from $\sim$ 3.4 to $\sim$3~GeV. The cutoff energy for tau
production is equal to
$$
\frac{m(m+2M)}{2(E_F+k_F)}.
$$
Then we enrich the model by using the effective mass. As a result
the cutoff energy is shifted to $\sim$3.2~GeV  and the total cross
section further reduced by about 7\%. The we take into account the
RPA corrections which results in a slight reduction of the total
cross section noticeable mainly for higher energies.

\begin{figure}[p]
\centerline{
\includegraphics[width=13cm]{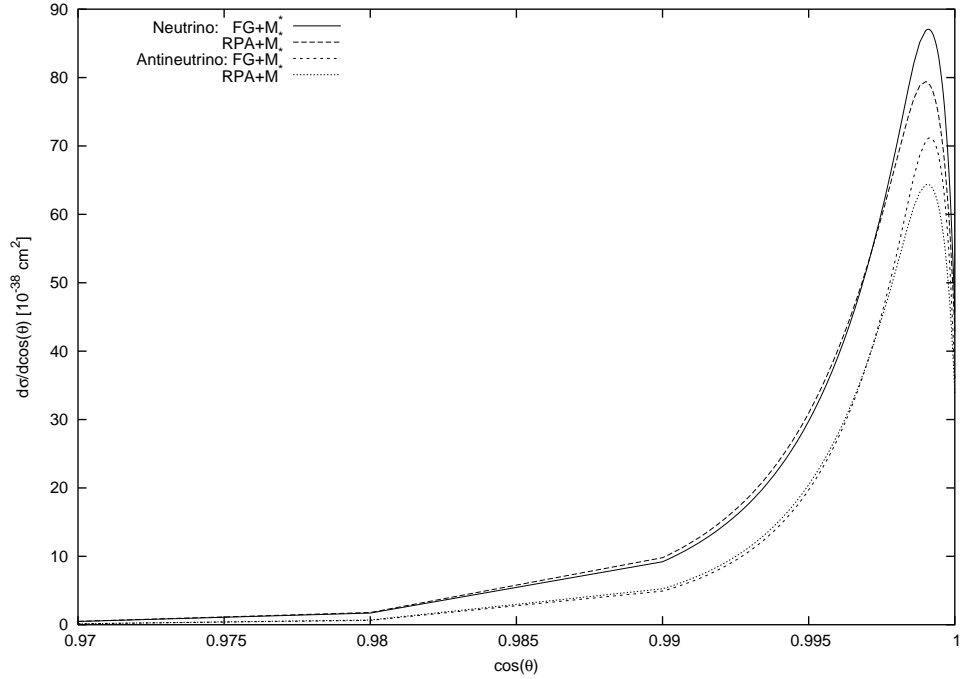}
} \caption{Comparison of the differential cross sections for
charged-current quasielastic neutrino(antineutrino)-nucleus
scattering. The nucleus is described by the Fermi gas with the
effective mass -- FG+$M^*$ (neutrino-nucleus scattering -- solid
line and antineutrino-nucleus -- short dashed line), and by the
Fermi gas with the effective mass and the RPA corrections --
RPA+$M^*$ (neutrino-nucleus scattering -- dashed line,
antineutrino-nucleus scattering -- dotted line). Calculations are
done for neutrino(antineutrino) energy of 7~GeV.
\label{wykres_pol2} }
\end{figure}

Fig.~\ref{wykres_pol2} illustrates how the introduction of the RPA
corrections influences the differential cross section obtained
within the Fermi gas model with the effective mass for neutrino
(antineutrino) energy of 7~GeV. The main contribution is located
in the region of small scattering angles where the RPA corrections
reduce the peak of the differential cross section by several
percent.

\begin{figure}
\centerline{
\includegraphics[width=14.5cm]{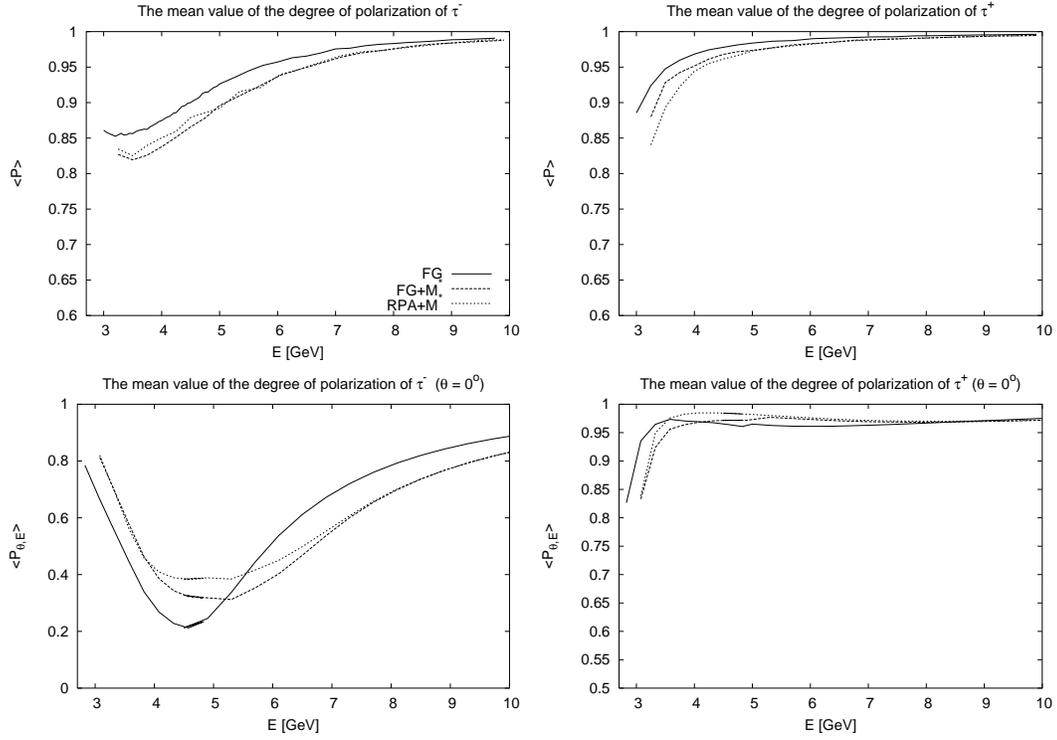}
} \caption{In the first row (Figs. a,b) the mean values (given by
equation (\ref{mean_P})) of the degree of polarization of
$\tau^\mp$ leptons are presented. The charts in the second row
(Figs. c, d) present the mean values (given by equation
(\ref{mean_P_E_theta})) of the degree of polarization of
$\tau^\mp$ leptons calculated for zero scattering angle. The tau
leptons are produced in charged-current quasielastic
neutrino(antineutrino)-nucleus scattering. The nucleus is
described by the Fermi gas (solid line -- FG), the Fermi gas with
the effective mass $M^*=$638~MeV (dashed line -- FG+$M^*$) and the
Fermi gas with the effective mass $M^*=$638~MeV and the RPA
corrections (dotted line -- RPA+$M^*$). \label{wykres_pol3} }
\end{figure}

The mean value of the degree of polarization of tau is defined as
\begin{equation}
\label{mean_P} \left<\mathcal{P}\right>=\frac{1}{\sigma(E)}\int d
E_\tau d \theta d\sigma(E_\tau,\theta,E)
\mathcal{P}(E_\tau,\theta,E).
\end{equation}
Its dependence on the neutrino (antineutrino) energy is shown in
Figs. \ref{wykres_pol3}(a) and \ref{wykres_pol3}(b) for three
cases: the  Fermi gas, the Fermi gas with the effective mass, and
the Fermi gas with the effective mass and the RPA corrections.

In the case of $\tau^-$ the mean value of degree of polarization
is almost constant ($\left<\mathcal{P}\right> \sim 0.85$ without
and about $0.82$ with the effective mass) up to 4~GeV then it
gradually rises to saturate at about 9~GeV. In the case of
$\tau^+$ the mean value of degree of polarization raises more
rapidly  saturating already at about 6~GeV. In both cases the
application of the effective mass to the free FG model decreases
$\left<\mathcal{P}\right>$ by a few percent and shifts the plots
to the right. It is interesting that the result of introducing the
RPA corrections into the Fermi gas with effective mass  is
opposite in the case of $\tau^+$ where it lowers the
$\left<\mathcal{P}\right>$ than in the case of $\tau^-$ where we
observe a slight enhancement.

The effects of the introduction of the effective mass and then the
RPA corrections are most visible in the Figs.~\ref{wykres_pol3}(c)
and \ref{wykres_pol3}(d) where  the mean value of the degree of
polarization of $\tau^-$ at zero scattering angle is plotted.
\begin{equation}
\label{mean_P_E_theta}
\left<\mathcal{P}_{\theta,E}\right>=\frac{1}{d\sigma(E,\theta)}\int
d E_\tau d\sigma(E_\tau,\theta,E) \mathcal{P}(E_\tau,\theta,E).
\end{equation}
The minimal value of the degree of polarization is obtained for
neutrino energy of about 4.5~GeV.  It is equal to 0.2 for the
Fermi gas model,  to 0.3 after the effective mass is used, and to
0.4 when the RPA corrections are also included.

\begin{figure}
\centerline{
\includegraphics[width=15cm]{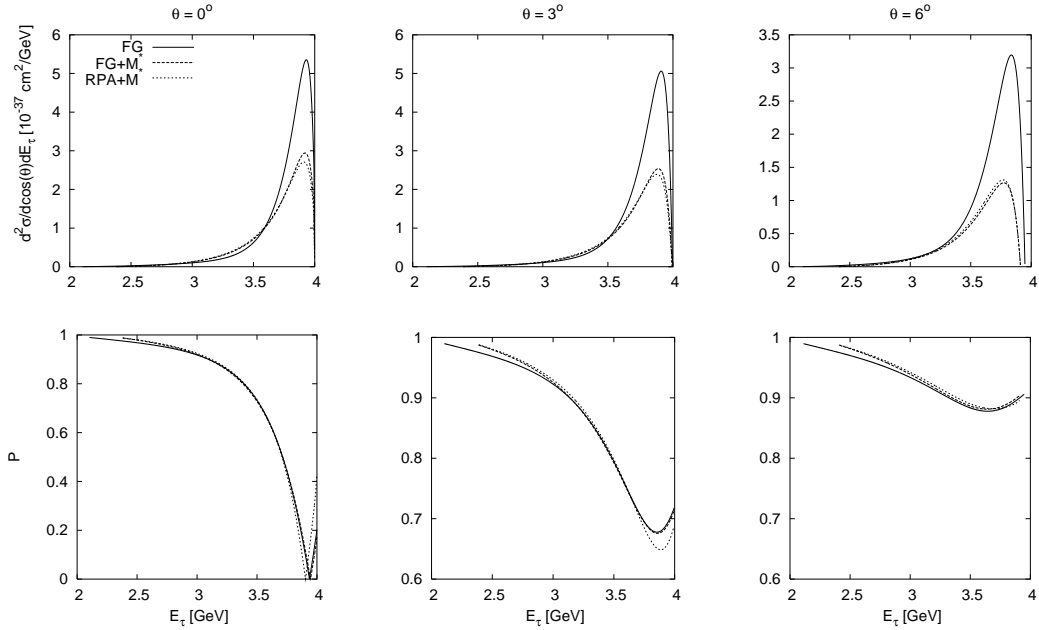}
} \caption{The charts in the first row present the differential
cross sections  calculated for three scattering angles: $\theta
=0^o,3^o,6^o $. The second row shows dependence of the degree of
polarization of the tau on its energy. The tau lepton is produced
in charged-current neutrino-nucleus scattering (for neutrino
energy 4~GeV). The nucleus is described by the Fermi gas (solid
line -- FG), the Fermi gas with the effective mass  $M^*=638$~MeV
(dashed line -- FG+$M^*$) and the Fermi gas with the effective
mass $M^*=638$~MeV and the RPA corrections (dotted line --
RPA+$M^*$). \label{wykres_pol5} }
\end{figure}

\begin{figure}
\centerline{
\includegraphics[width=15cm]{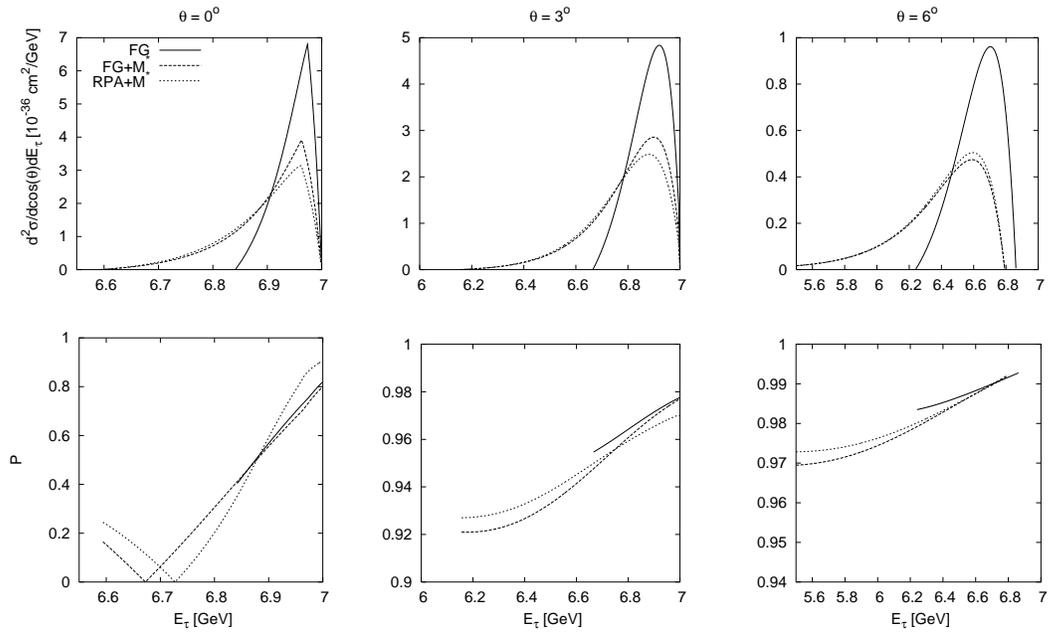}
} \caption{ The same as Fig.~\ref{wykres_pol5} but neutrino energy
is 7~GeV. \label{wykres_pol6}}
\end{figure}

\begin{figure}
\centerline{
\includegraphics[width=15cm]{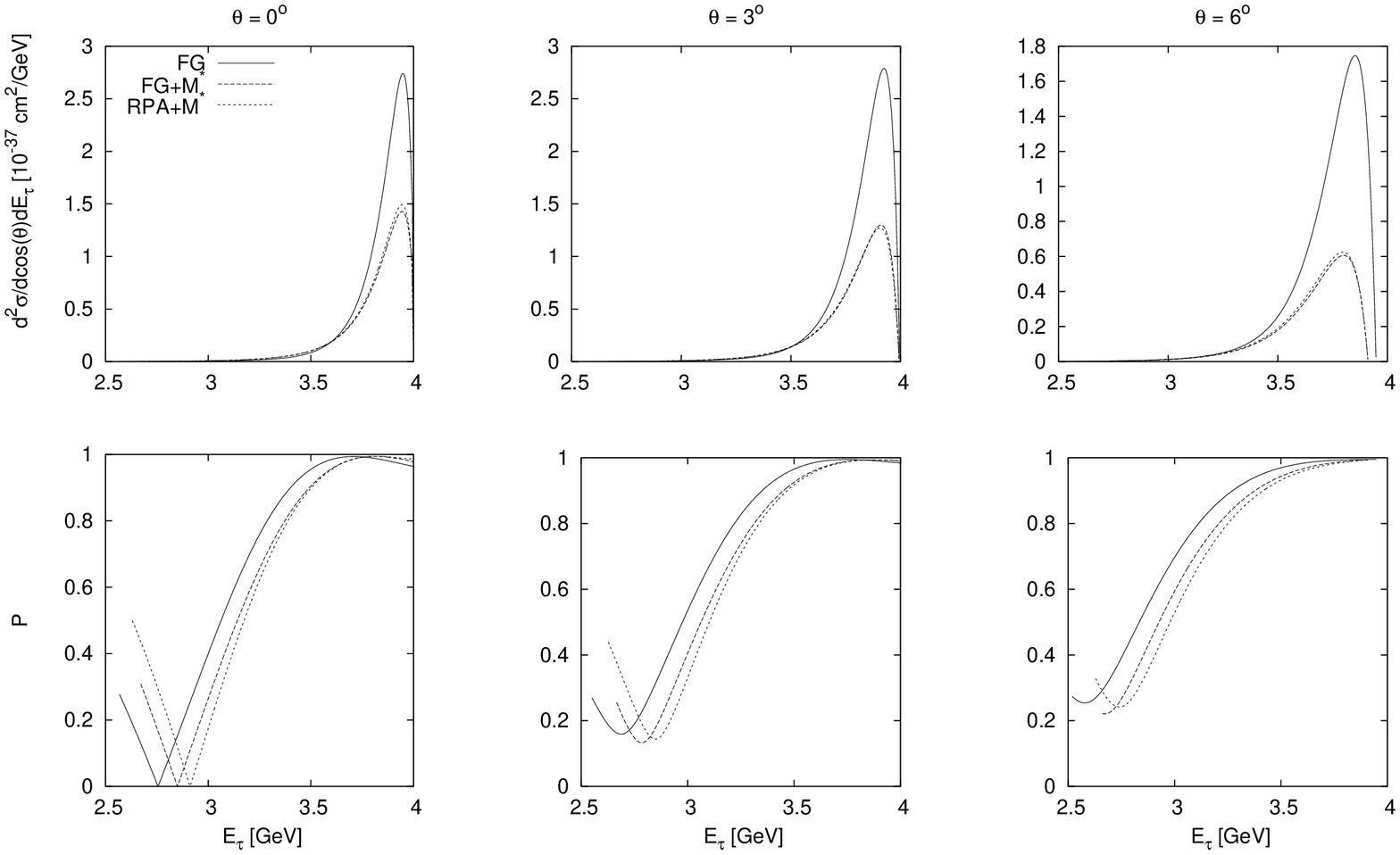}
} \caption{The same as Fig.~\ref{wykres_pol5} but for
antineutrino.\label{wykres_pol7} }
\end{figure}

\begin{figure}
\centerline{
\includegraphics[width=15cm]{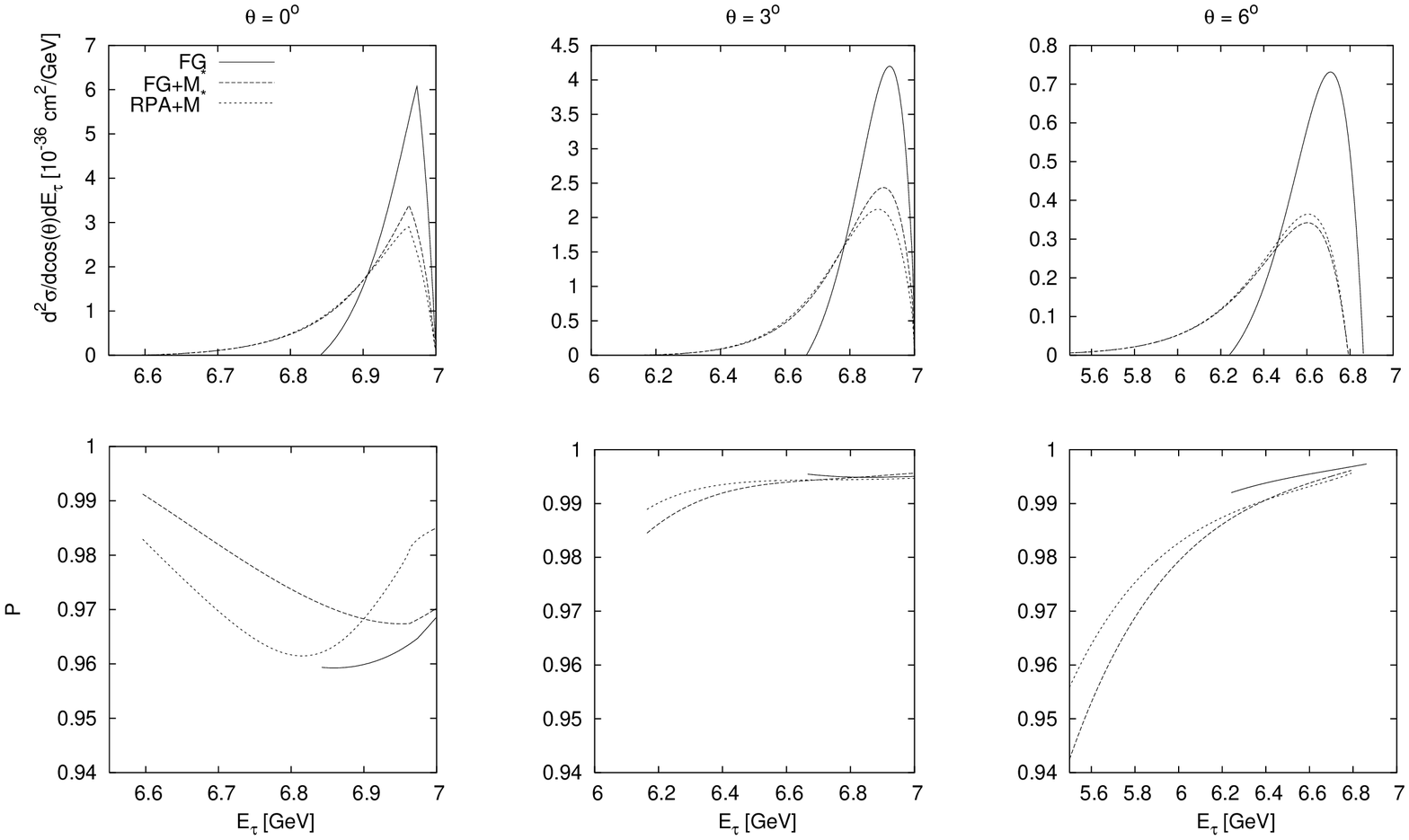}
} \caption{ The same as Fig.~\ref{wykres_pol6} but for
antineutrino.\label{wykres_pol8} }
\end{figure}

In Figs.~\ref{wykres_pol5},~\ref{wykres_pol6},
\ref{wykres_pol7},~\ref{wykres_pol8} the dependence of the degree
of polarization on the  tau energy is presented. Calculations are
done for three scattering angles ($0^o$, $3^o$, $6^o$). We compare
the plots of polarization  with the appropriate differential cross
sections given by equation (\ref{przekroj}).

In general, for a given scattering angle, there exist two
kinematically allowed regions for tau energy. One is placed close
to the tau mass, and the other  is placed near the
neutrino(antineutrino) energy. The Fermi motion widens these
regions and for the beam energy of 4~GeV they  join each other.
For higher neutrino (antineutrino) energies, these two regions are
separated by a large forbidden area. The plots of $\mathcal{P}$
and $d^2\sigma
 / d \cos(\theta)dE_\tau $ for E= 7~GeV presented in
 Figs.~\ref{wykres_pol6},~\ref{wykres_pol8}
 are confined to the area of
energy close to the neutrino (antineutrino) energy because in the
other area the cross section is negligible.

One can notice that the degree of polarization strongly depends on
the scattering angles and the produced leptons have high degree of
polarization for almost  all angles, apart from $\theta=0$, where
the tau degree of polarization can be small.

Figs.~\ref{wykres_pol5} and \ref{wykres_pol7} show the results for
scattering of 4~GeV neutrinos and antineutrinos. It can be seen
that the polarization plots for $\theta=0^o$ have sharp minimum
where $\mathcal{P}$ reaches zero. In this point the longitudinal
polarization changes its sign from positive to negative. This
effect is clearly explained in~\cite{japonczycy}. The helicity of
$\tau^-$ in the LAB frame changes from right-handed to left-handed
(for $\tau^+$ there is an analogical effect). For higher angles we
can still see a local minimum in this point but it is much higher
and is smooth because of the non zero contribution from the
perpendicular polarization. It is interesting that this point is
shifted after the introduction of the effective mass and also
after the inclusion of the RPA corrections (it can be clearly seen
in Figs.~\ref{wykres_pol6} and \ref{wykres_pol7}).

For the neutrino(antineutrino) energy of 7~GeV the two
kinematically allowed regions correspond to opposite signs of the
longitudinal polarization of the tau (the minimum region is
kinematically forbidden) when $M^*=M$. However, when
$M^*=638$~MeV, it is possible that both signs of $\mathcal{P}_l$
may be present in the same kinematical region which makes  the
minimal value of $\mathcal{P}=0$ available for zero scattering
angle. Lowering of the effective mass stretches the allowed
kinematical regions reducing the gap between them.

Finally, we apply the Fermi gas model with the RPA corrections and
the LDA to the case of the ICARUS detector and the CNGS beam. We
fix the tau neutrinos energy at 7~GeV -- the most probable energy
of the tau neutrinos resulting from the oscillations  of the CNGS
beam muon neutrinos.

\begin{figure}
\centerline{
\includegraphics[width=15cm]{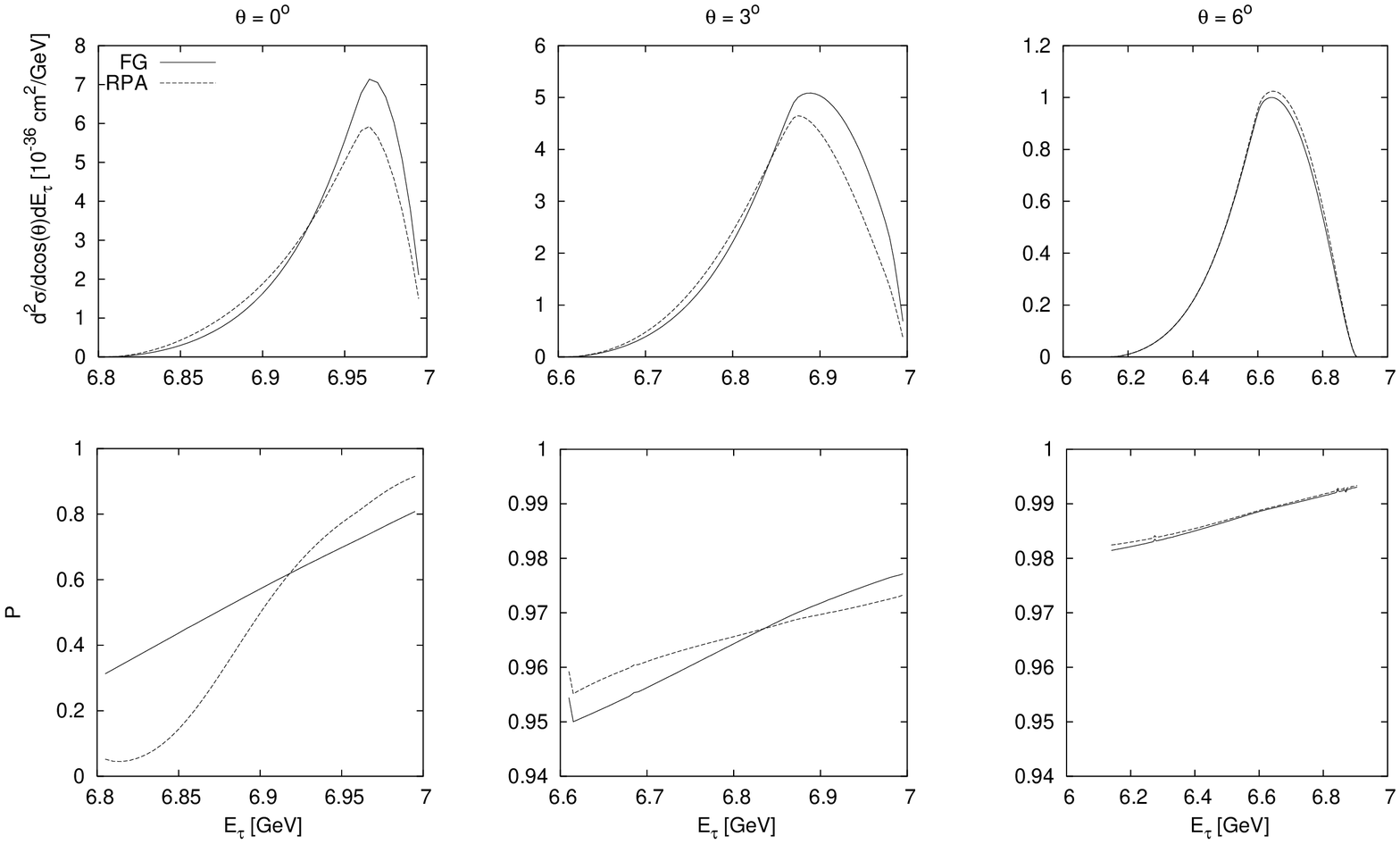}
} \caption{ The charts in the first row present the differential
cross sections  calculated for three scattering angles: $\theta
=0^o,3^o,6^o $. The second row shows dependence of the degree of
polarization of the tau on its energy. The tau lepton is produced
in charged-current neutrino-nucleus scattering (for neutrino
energy 7~GeV). The nucleus is described using the LDA approach for
argon and by the Fermi gas (solid line -- FG)  and the Fermi gas
with the RPA corrections (dotted line --
FG+RPA).\label{wykres_pol10} }
\end{figure}

In the description of the argon nucleus (the target in the ICARUS
detector) we use the LDA approach. We apply the charge density
profile from atomic data tables~\cite{DeVries} to describe nucleon
density $\rho(r)$ in nucleus  (see appendix). The Fermi momenta
for protons and neutrons are the following:
\begin{eqnarray}
\label{sredni_kf}
{k_F}_{p}(r) & = & \sqrt[3]{3\pi^2\rho_p(r)},\quad \rho_p(r)=
\textstyle \frac{Z}{A}\rho(r),\\
{k_F}_{n}(r) & =
&\sqrt[3]{3\pi^2\rho_n(r)},\quad\rho_n(r)=\textstyle
\frac{A-Z}{A}\rho(r).
\end{eqnarray}
where the atomic number $A=\int d^3 r \rho(r)=40$, and $Z=18$. The
cross section is given by the integral:
\begin{equation}
d\sigma(E_\tau,\theta) =  \frac{1}{A}\int d^3 r \rho_{n,p}(r)
d\sigma({k_F}_{n,p}(r),E_\tau,\theta).
\end{equation}
The results are shown in Fig.~\ref{wykres_pol10} where we compare
the degree of polarization of tau and the appropriate differential
cross section obtained with and without the RPA corrections. As in
the case of the global Fermi momentum (see Fig.~\ref{wykres_pol6})
the introduction of the RPA is the most visible for the zero
scattering angle. The plot of the degree polarization (in the case
of the FG) is a straight line rising from 0.4 to 0.8. The
application the RPA corrections  yields much deeper minimum (less
than 0.1) and a maximum reaching 0.9.

\vskip 0.5cm \textbf{Summary} \vskip 0.1cm

The tau leptons produced in neutrino, antineutrino-nucleus
scattering are characterized by  high degree of polarization.
Nevertheless, the $\tau^-$ can be almost unpolarized at zero
scattering angle for small neutrino energies. In particular, for
the neutrino energy of 4.5~GeV, the mean degree of polarization
reaches its minimum. The minimal value of
$\left<P_{\theta,E}\right>$ is very sensitive to the details of
the model such as the effective mass and the RPA corrections.
Introducing the effective mass increases it by about 50\% (from
about 0.2 to 0.3) and including the RPA corrections causes and
additional 33\% rise (from 0.3 to 0.4).

It should be also mentioned that the RPA corrections influence the
energy for which the tau's helicity changes the sign, whereas the
introduction of the effective mass strongly affects the kinematics
by broadening the kinematically allowed regions and causing the
gap between them to disappear for some neutrino energy values.

The results of this paper are quite similar to those obtained for
the neutrino-nucleon deep inelastic scattering in
  \cite{japonczycy,Graczyk_pol}. It suggests that it is the
kinematics that plays the main role in the calculations of the
polarizations.

\vskip 0.5cm

\textbf{Acknowledgments}

\vskip 0.5cm

I would like to thank  J. Sobczyk for stimulating discussions and
useful remarks which helped to improve this paper. I thank D.
Kie\l\-czew\-ska and  E. Rondio, as well as K. Kurek for
interesting discussions. I also would like to thank C. Juszczak
for reading the manuscript and interesting comments.

 \vskip 0.5cm
 This work was supported by the KBN grant 105/E-344/SPB/ICARUS/P-03/DZ
211/2003-2005.

The author is a Max Born Scholarship fellow.

\appendix

\section{Basis Vectors }
The four-momenta of the neutrino and the tau have the form:
$$k_\mu=(E, k_1, k_2, k_3)$$
and
$$
{k'}_\mu = (E-q_0, k_1 - q, k_2, k_3).
$$
Appropriate calculations lead to the expressions for basis
vectors:
\begin{eqnarray*}
e^l_\mu & = & \frac{\chi
E_\tau}{m}\left(\frac{|\mathbf{k'}|}{E_\tau}, \frac{
k_1-q}{|\mathbf{k'}|},
\frac{k_2}{|\mathbf{k'}|}, \frac{ k_3}{|\mathbf{k'}|}\right),\\
e^t_\mu & = & \frac{1}{\sqrt{k_2^2 + k_3^2}}(0, 0, -k_3, k_2),\\
e^p_\mu & = & -\frac{1}{|\mathbf{k'}|\sqrt{k_2^2 +k_3^2}}(0,k_2^2
+ k_3^2, (k_1-q)k_2, (k_1-q)k_3),
\end{eqnarray*}
where $\chi=\mp1$ describes helicity of $\tau^{\mp}$.

\section{Density profile of Argon}

Charge density profile of $_{18} Ar^{40}$ is the following
\cite{DeVries}:
\begin{equation}
\label{density_ArFe} \rho(r)=\frac{\rho(0)}{\left(
1+\exp\left((r-C)/C_1\right)\right)},
\end{equation}
$$\rho(0)=0.176\,\rm{fm}^{-3},\;\;\; C=3.530\,\rm{fm},\;\;\;
C_1=0.542\,\rm{fm}.$$


\end{document}